\documentstyle[aps,prl,epsfig,multicol]{revtex}
\begin{document}
\draft
\title{Unusual dependence of vortex core states on the
superconducting gap \\ in Bi$_2$Sr$_2$CaCu$_2$O$_{8+\delta}$}
\author{B.~W.~Hoogenboom,$^{1,}$\cite{email} K.~Kadowaki,$^2$
B.~Revaz,$^{1,}$\cite{addressRevaz} M.~Li,$^3$ Ch.~Renner,$^4$ and
\O.~Fischer$^1$}
\address{$^1$ DPMC, Universit\'e de Gen\`eve, 24 Quai Ernest-Ansermet,
1211 Gen\`eve 4, Switzerland}
\address{$^2$ Institute of Materials Science, University of Tsukuba,
Tsukuba, Ibaraki 305-8573, Japan}
\address{$^3$ Kamerlingh Onnes Laboratory,
Leiden University, P.O. Box 9506, 2300 RA Leiden, The Netherlands}
\address{$^4$ NEC Research Institute, 4 Independence Way, Princeton,
New Jersey 08540, USA}
\date{\today}
\maketitle
\begin{abstract}
We present a scanning tunneling spectroscopy study on
quasiparticle states in vortex cores in
Bi$_2$Sr$_2$CaCu$_2$O$_{8+\delta}$. The energy of the observed
vortex core states shows an approximately linear scaling with the
superconducting gap in the region just outside the core. This
clearly distinguishes them from conventional localized core
states, and is a signature of the mechanism responsible for their
discrete appearance in high-temperature superconductors. The
energy scaling of the vortex core states also suggests a common
nature of vortex cores in Bi$_2$Sr$_2$CaCu$_2$O$_{8+\delta}$ and
YBa$_2$Cu$_3$O$_{7-\delta}$. Finally, the observed vortex core
states do not show any dependence on the applied magnetic field in
the range from 1 to 6~T.
\end{abstract}

\begin{multicols}{2}
\narrowtext


In conventional, $s$-wave superconductors, the suppression of the
order parameter in a vortex core creates a potential well for
low-energy quasiparticles, leading to the formation of localized
states \cite{Caroli:1964,Hess:1989,Renner:1991}. On the contrary,
if the superconducting order parameter has nodes in it -- as for
$d_{x^2-y^2}$ symmetry in high-temperature superconductors (HTS)
-- one expects the low-energy quasiparticles states in a vortex to
be extended along the nodes of the gap function, so to be
delocalized. This would result in a broad peak at the Fermi level
in the quasiparticle local density of states (LDOS) of a vortex
core \cite{Wang:1995,Franz:1998,Yasui:1999}. In this light, the
observation of {\em discrete} vortex core states in
YBa$_2$Cu$_3$O$_{7-\delta}$ (YBCO) \cite{Maggio:1995} and
Bi$_2$Sr$_2$CaCu$_2$O$_{8+\delta}$ (BSCCO)
\cite{Hoogenboom:2000,Pan:2000a} by scanning tunneling
spectroscopy (STS) has come as a complete surprise. As a result,
the nature of these states has been subject of increasing
theoretical study
\cite{Wang:1995,Franz:1998,Yasui:1999,Volovik:1993,Soininen:1994,Morita:1997,Himeda:1997,Franz:2000,Han:2000,Balatsky:2000},
leading to a range of possible scenarios for explaining the
experimental data. The need for understanding of the electronic
structure of the vortices in HTSs has become even more pressing
because of the antiferromagnetic fluctuations recently observed in
vortex cores in La$_{2-x}$Sr$_x$CuO$_4$ \cite{Lake:2001}.

To our knowledge, the most direct way to access the electronic
structure of a vortex core is by using a scanning tunneling
microscope (STM). In a typical experimental set-up for studying
vortex cores in HTSs (as well as in this study), the STM tip and
tunneling direction are perpendicular to the CuO$_2$ planes. The
quasiparticle excitation spectrum then follows from the d$I$/d$V$
tunneling spectra \cite{Wolf:1985}. Experiments on HTSs have
revealed the following characteristics of vortex cores. In YBCO
the vortex core states appear as two clearly distinct, sub-gap
excitations, which do not disperse on moving out of the vortex
core, but rather transform into weak shoulders in the
superconducting spectra (and sometimes also observed at zero
magnetic field) \cite{Maggio:1995}. In BSCCO the vortex core
spectra reveal a remarkable resemblance to the pseudogap spectra
observed above the critical temperature $T_c$ \cite{Renner:1998}.
In addition, vortex core states appear as weak shoulders in this
pseudogap, do not change in energy as a function of increasing
distance from the vortex core \cite{Hoogenboom:2000} (contrary to
the vortex core states in conventional superconductors
\cite{Hess:1989}), and decay over a characteristic length scale of
about 20~\AA\ \cite{Pan:2000a}. Though quite irregular shapes can
be observed due to vortex motion \cite{Hoogenboom:2000a}, the
cores do no show any sign of the four-fold symmetry that may be
expected for $d$-wave superconductors.

In this letter, we present new data on vortex cores in BSCCO,
obtained with a low-temperature STM. Our main results can be
summarized as follows. First, the energy of the vortex core states
scales with the superconducting gap outside the core, clearly
distinguishing these states from localized vortex states
($E\propto\Delta^2/E_F$) in conventional superconductors
\cite{Caroli:1964}. Since in BSCCO the superconducting gap scales
with the oxygen doping level
\cite{Oda:1997,Renner:1998a,DeWilde:1998,Miyakawa:1998,Miyakawa:1999,Matsuda:1999},
this directly gives the doping dependence of the vortex states as
well. Second, the vortex core spectra do not show any significant
dependence on the external magnetic field over a range from 1 to
6~T, which questions field-dependent scenarios for explaining the
vortex core states.

A systematic study of vortex core spectra requires, apart from
sufficient instrumental resolution, relatively high sample
homogeneity. More precisely, in order to compare spectra inside a
vortex core to those in the nearby superconducting region,
(zero-field) spectral reproducibility over at least 100~\AA\ is
necessary. This condition can be met in BSCCO single crystals with
transition widths $\Delta T_c \leq 1$~K (as measured by AC
susceptibility). Most of the data presented here were obtained,
with magnetic field and tunneling direction perpendicular to the
CuO$_2$ planes, on two different overdoped samples: (i) A sample
with $T_c=77.7$~K ($\Delta T_c=0.4$~K), cooled down at 6~T, and
after measurements at 6~T the field was reduced (at low
temperature) to 4 and 1~T. (ii) A sample with $T_c=77$~K ($\Delta
T_c=1$~K), zero-field cooled, measured at fields of subsequently
0, 6, 2, and 0~T. For the latter sample, lines of spectra in and
near vortex cores, as well as vortex images as a function of field
can be found in Refs.~\cite{Hoogenboom:2000,Hoogenboom:2000a}.
Some additional data were taken on two other overdoped samples
with $T_c=76.1$~K ($\Delta T_c=0.3$~K), and on an optimally doped
sample with $T_c=87.4$~K ($\Delta T_c=1.0$~K). Samples were
cleaved in ultra-high vacuum environment ($10^{-9}$~mbar), shortly
before cooling down STM and sample to 4.2~K in exchange gas ($\sim
10^{-2}$~mbar helium). All measurements in this study were taken
at 4.2~K. For further experimental details we refer to
Refs.~\cite{Renner:1990,Kent:1992,Renner:1995}.

The quasiparticle LDOS at an energy $E=eV$ on the surface of the
sample was obtained by measuring the differential tunneling
conductance d$I$/d$V$ at sample bias $V$ as a function of
position, using a lock-in technique. In homogeneous BSCCO samples,
clearest vortex images were found by measuring the conductance at
$V=-\Delta_p/e$, where $\Delta_p$ corresponded to the energy of
the superconducting coherence peaks. We thus obtained maps of
spots where superconductivity was suppressed. A comparison of the
number of these spots per area to the magnetic flux for different
fields, as well as their disappearance at zero field, justified
the identification as vortex cores
\cite{Renner:1998,Hoogenboom:2000a}.

\begin{figure}
 \epsfysize=40mm
 \centerline{\epsffile{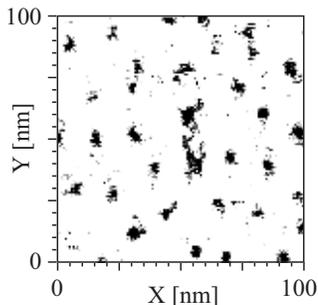}}
 \caption{Image of vortex cores at 6~T (field cooled measurement),
 taken by measuring the conductivity d$I$/d$V$ at -30~mV, and
 normalized on its value at zero bias.}
 \label{map}
\end{figure}

\begin{figure}
 \epsfxsize=70mm
 \centerline{\epsffile{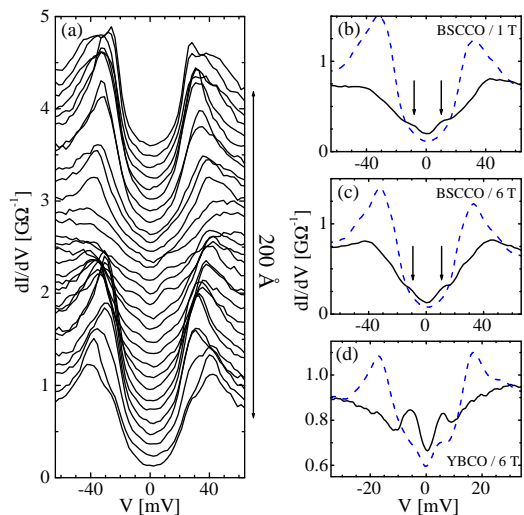}}
 \caption{
 (a) Spectra along a 200~\AA \ trace through a vortex core in
 BSCCO at 1~T. The vortex core is about halfway the trace. The
 spectra have been offset for clarity.
 (b) Averaged spectrum in the centre of this core, and in its
 immediate vicinity (dashed line). The latter spectrum may appear a
 bit broad, since it is the average of several spectra with
 peaks at slightly different energies.
 (c) For a vortex core in BSCCO, at 6~T.
 (d) For a  vortex core in YBCO at 6~T. Note the different scales
 for the bias voltage.}
 \label{spectra}
\end{figure}

In Fig.~\ref{map} we show a map of vortex cores at 6~T. As can be
verified directly, the number of vortices ($30\pm2$) in
Fig.~\ref{map} corresponds to the average flux crossing this area
($29\Phi_0$) at 6~T. In general, the density of vortices scales
with the magnetic field as one should expect. We do not observe
any systematic change in the size and shape of the vortex cores
for the different applied fields.

d$I$/d$V$ spectra were taken in and around several of the thus
imaged vortex cores. Spectra along a 200~\AA \ trace through a
vortex core are shown in Fig.~\ref{spectra}(a). The spectra
discussed hereafter were obtained by averaging the spectra just
outside the core [above and below in Fig.~\ref{spectra}(a)] and
the spectra well inside the core [in the middle of
Fig.~\ref{spectra}(a)], for the superconducting spectra and the
typical vortex core spectra respectively. These averaged spectra
can be found in Fig.~\ref{spectra}(b-d), for a vortex core at 1~T
and for one at 6~T in BSCCO, as well as for a vortex core in YBCO
\cite{Maggio:1995}). Compared to the superconducting state, the
LDOS in the vortex cores seems considerably reduced, with loss of
spectral weight near the Fermi level. The characteristics of the
vortex spectra in BSCCO, asymmetric pseudogap with weak shoulders
at low bias, do not show any dependence on the field strength. We
neither observe, within the experimental resolution, any change
(due to the magnetic field) in the spectra outside the vortex
core. The spectra taken at different field strengths are in fact
remarkably similar, both those inside, and those around vortices.

In order to study the dependence of the vortex core spectra on the
superconducting energy gap (taken as the energy of the coherence
peaks $\Delta_p$), we measured vortex cores on different samples,
and also made use of a certain inhomogeneity (on a scale larger
than 100~\AA) of some of the samples which we would characterize
as moderately homogeneous. (In fact large-scale homogeneity has
also been obtained, using an optimum oxygenation procedure.)
$\Delta_p$ has been determined from the average of spectra around
the vortex core, and its precision depends on the inhomogeneity of
the superconducting order parameter in immediate neighborhood of
the core.

\begin{figure}
 \epsfxsize=90mm
 \centerline{\epsffile{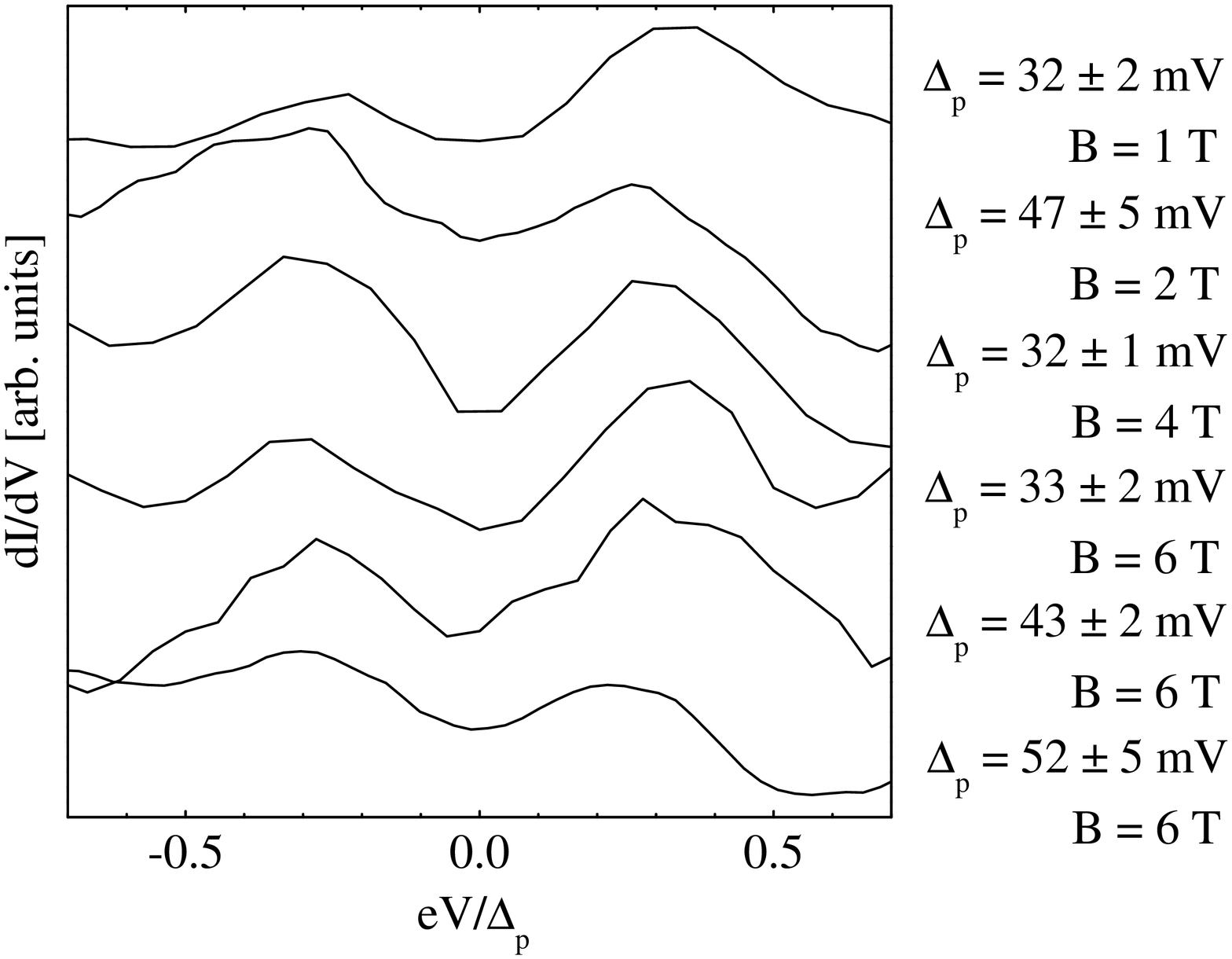}}
 \caption{Spectra of vortex core states obtained by subtracting a
 fitted background (see text) from the vortex core spectra,
 plotted with an energy scale normalized on the energy of the
 coherence peak $\Delta_p$ of spectra near the core. There is no
 significant variation in the spectra for different magnetic fields
 and $\Delta_p$.}
 \label{subtract}
\end{figure}

For a precise and consistent determination of the energies of the
core states, the vortex core spectra were fitted with a fifth
order polynomial over a range $0.6\Delta_p<eV<1.5\Delta_p$, with
the fits forced to go through the zero-bias conductance. These
fits of the pseudogap background were subtracted from the vortex
spectra, thus leaving the excess spectral weight related to vortex
core states. For the vast majority of vortices, the core state
energies as determined from the fitted and subtracted data fall
within the error bars of the core state energies determined
directly from the raw data (see the arrows in Fig.~\ref{spectra}).
The latter are less precise because the core states are partly
hidden in the slope of the pseudogap \cite{Hoogenboom:2000}. The
results have been checked for robustness against variation of the
fitting parameters and range, and have been plotted on a
normalized energy scale in Fig.~\ref{subtract}.

\begin{figure}
 \epsfxsize=75mm
 \centerline{\epsffile{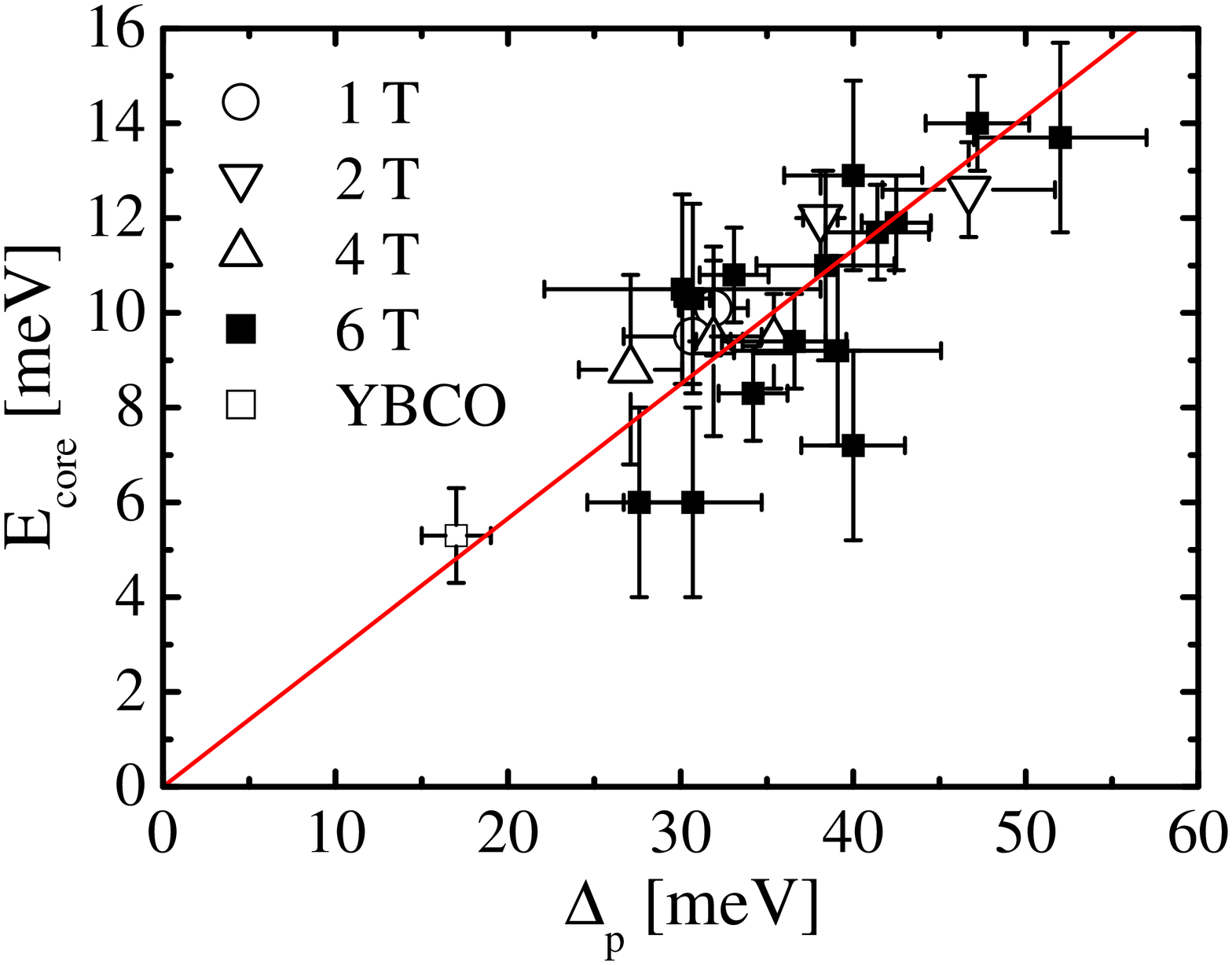}}
 \caption{The energy of vortex core states $E_{core}$ as a function
 of the superconducting gap (the energy of the coherence peaks)
 $\Delta_p$. Plotted are data for different magnetic fields, and
 different BSCCO single crystals. At $\Delta_p=17$~meV the value
 extracted from YBCO data. The straight line is a linear fit to all
 data points, forced to go through the origin, with a slope of
 0.28. An unrestricted linear fit would cut the vertical axis at
 about 1~meV, the latter value also being the uncertainty of the
 offset.}
 \label{straight}
\end{figure}

These data confirm the independence of the vortex states on the
magnetic field, and also show that their energy $E_{core}$
directly scales with the superconducting gap. This scaling comes
out even more clearly when $E_{core}$ is plotted as a function of
$\Delta_p$, for all measured vortices in different samples and at
different fields, see Fig.~\ref{straight}. The dependence on
$\Delta_p$ is certainly not quadratic, as one would expect for
conventional localized states in vortex cores \cite{Caroli:1964}.
It is roughly linear, passing through the origin, and with a slope
of 0.28. We have checked that the energies extracted from the raw
data, with a bit more scattering of the data and a slope of
approximately 0.30, do show the same behavior. This linear
behavior is an important result, since it proves that the vortex
core states in BSCCO do not correspond to conventional localized
states.

Since doping dependent tunneling experiments on BSCCO show a
roughly linear dependence of $\Delta_p$ on oxygen doping
\cite{Oda:1997,Renner:1998a,DeWilde:1998,Miyakawa:1998,Miyakawa:1999,Matsuda:1999},
we can directly read the horizontal scale as the (local) hole
doping of the BSCCO samples, going from overdoped to underdoped.
This indicates that the electronic nature of the vortex cores
remains the same for a considerable doping range.

In Fig.~\ref{straight} we have also included data obtained on
YBCO, for comparison \cite{Maggio:1995}. Though the vortex core
states in YBCO appear much more pronounced than those in BSCCO
(see Fig.~\ref{spectra}), their energy scale shows a remarkable
similarity, strongly suggesting a common origin. It is tempting,
but rather speculative, to compare this to the energy scale of the
antiferromagnetic fluctuations with an energy of $3\sim4$~meV, for
a spin gap of 6.7~meV, as observed in La$_{2-x}$Sr$_x$CuO$_4$
\cite{Lake:2001}.

Our results, especially the linear scaling of $E_{core}$ with
$\Delta_p$, rule out conventional localized states as an
explanation for the vortex core states observed in HTSs (they
would neither be consistent with the lack of dispersion as a
function of position, see Fig.~\ref{spectra}). Furthermore, they
provide sufficient details to discuss other scenarios.

First of all, in a pure $d$-wave superconductor the vortex core
states would appear as a broad zero-bias conductance peak
\cite{Wang:1995,Franz:1998,Yasui:1999}, in clear disagreement with
experiment. One may avoid this difficulty by supposing that the
quasiparticle states from different vortex cores form bands, which
would lead to the splitting of the zero-bias conductance peak into
two different peaks. In that case, however, the energy gap between
these peaks should depend on the magnetic field, which is
inconsistent with the results presented above.

A possible explanation for the observed vortex core states is the
symmetry breaking of the order parameter in the vicinity of a
vortex core \cite{Volovik:1993,Soininen:1994}. This will lead to
secondary (possibly imaginary) $d_{xy}$ or $s$ components of the
order parameter, effectively blocking the nodes of the
$d_{x^2-y^2}$ gap function, and allowing localized states
\cite{Franz:1998}. Though in a simple BCS $d$-wave superconductor
these components are too small to influence the spectra
\cite{Yasui:1999}, they appear more important when the Coulomb
interaction on the Cu sites is taken into account. Numerical
calculations using the $t-J$ model estimate their size -- strongly
doping dependent -- to be between 5 and 30\% of the
$\Delta_{x^2-y^2}$ order parameter \cite{Himeda:1997,Han:2000}.
Our results imply that such a secondary component would scale with
the $\Delta_{x^2-y^2}$ gap over a wide doping range and, contrary
to what one should expect \cite{Yasui:1999,Balatsky:2000}, would
be independent of the magnetic field between 1 and 6~T.

In this context, one may ask whether the observed core states are
directly related to the vortices, or are merely {\em enhanced}
inside the cores. In the secondary order parameter scheme, the
latter possibility would mean that this secondary component is
also present at zero field. In fact, low-energy features similar
to vortex core states (though considerably less pronounced) have
been observed in the gap of YBCO at zero field \cite{Maggio:1995},
and interpreted as the signature of an additional $s$ or $d_{xy}$
component in the order parameter in overdoped YBCO
\cite{Yeh:2001}.

The energy scaling and field independence of the core states
naturally suggest an other scenario, involving the pseudogap
state. Keeping in mind that the pseudogap above $T_c$ (and in the
vortex cores) scales with the superconducting gap as well
\cite{Renner:1998a,Miyakawa:1999,Matsuda:1999,Kugler:2001}, one
can speculate that the presence of the pseudogap leads to a
splitting of the BCS $d$-wave zero bias conductance peak. This
idea has been elaborated in two very recent calculations, using
the two-body Cooperon operator for modeling phase fluctuations
\cite{Berthod:2001} and invoking spin-density wave order in the
vortex cores \cite{Zhu:2001}. The pseudogap being much more
dominant in BSCCO than in YBCO, it will suppress the low-energy
core states in the former material much more than in the latter
\cite{Berthod:2001}, in agreement with our observations
(Fig.~\ref{spectra}).

In conclusion, we have measured vortex core quasiparticle states
as a function of the superconducting gap and the magnetic field.
Their energy dependence on the superconducting gap near the vortex
core is approximately linear, with a slope of about 0.28. The
vortex core states do not show any dependence on the applied
magnetic field in the range from 1 to 6~T. Our results suggest a
common origin for vortex core states in YBCO and BSCCO. The linear
energy scaling of these states with the superconducting gap, the
lack of any dispersion as a function of distance from the vortex
core centre, and the suppression of LDOS in the core compared to
the superconducting state further underline the non-BCS behavior
of HTSs, and of their vortex cores in particular.

We acknowledge C.~Berthod and B.~Giovannini for very useful
discussions, J.-Y.~Genoud for assistance in sample preparation,
and I.~Maggio-Aprile for providing the YBCO data.



\end{multicols}
\end{document}